# AI for Chemical Space Gap Filling and Novel Compound Generation

Monee Y. McGrady[1], Sean M. Colby[1], Jamie R Nuñez[1], Ryan S. Renslow[1, *], Thomas O. Metz[1, *]

[1] Biological Sciences Division, Pacific Northwest National Laboratory, Richland, WA, USA

* ryan.renslow@pnnl.gov
* thomas.metz@pnnl.gov

**ABSTRACT:** When considering large sets of molecules, it is helpful to place them in the context of a "chemical space" – a multidimensional space defined by a set of descriptors (e.g., molecular properties) that can be used to visualize and analyze compound grouping as well as identify regions that might be void of valid structures. The chemical space of all possible molecules in a given biological or environmental sample can be vast and largely unexplored, mainly due to current limitations in processing of 'big data' by brute force methods (e.g., enumeration of all possible compounds in a space). Recent advances in artificial intelligence (AI) have led to multiple new cheminformatics tools that incorporate AI techniques to characterize and learn the structure and properties of molecules in order to generate plausible compounds, thereby contributing to more accessible and explorable regions of chemical space without the need for brute force methods. We have used one such tool, a deep-learning software called DarkChem, which learns a representation of the molecular structure of compounds by compressing them into a latent space. With DarkChem's design, distance in this latent space is often associated with compound similarity, making sparse regions interesting targets for compound generation due to the possibility of generating novel compounds. In this study, we used 1 million small molecules (less than 1000 Da) to create a representative chemical space (defined by calculated molecular properties) of all small molecules. We identified regions with few or no compounds and investigated their location in DarkChem's latent space. From these spaces, we generated 694,645 valid molecules, all of which represent molecules not found in any chemical database to date. These molecules filled 50.8% of the probed empty spaces in molecular property space. Generated molecules are provided in the supporting information.

## INTRODUCTION

Chemical space is a term used to describe the realm of all possible molecules. Professor Jean-Louis Reymond (University of Bern), who leads a group well known for their revolutionary work in this field, describes chemical space as "a concept to organize molecular diversity by postulating that different molecules occupy different regions of a mathematical space where the position of each molecule is defined by its properties".[1] As the broadest representation of chemical space theoretically encompasses all possible molecules in our universe (estimated at $10^{180}$ structures for molecules with molecular weight under 1000 Da)[2], having an understanding of the organization and structure of this space could lead to new insights into yet undiscovered molecules and the relationships between them. For example, representations of chemical space are already being used to search areas around lead-like drug compounds for novel drug candidates (i.e., mining chemical spaces defined by chemical properties of known molecules to discover new, functionally important molecules).[3-5] It has been observed that, for chemical spaces generated using both experimentally observed and artificially/computationally generated molecules, there are regions in chemical space that remain sparse or have no representation.[6] It is implausible that all gaps could eventually be populated due to restrictions governed by the laws of physics (i.e., some combinations of chemical property values may be mutually exclusive). To date, an all-encompassing chemical space has yet to be defined, as the following remain unknown: 1) the minimum set of chemical properties required to fully define the dimensionality of such a space, 2) the chemical property value ranges that bound the size of the dimensions, and 3) the gaps in this space that will forever remain void due to physical law restrictions. However, even by creating naïve or incomplete representations of chemical space, we can still gain a better understanding of where current gaps in known chemical libraries are located and it is possible to begin filling them with compounds generated using various tools.

To date, multiple approaches have been used to define limited chemical spaces. Enumerative approaches, such as the multiple "Generated Database" (GDB) of molecules created by the Reymond group,[7-9] focus on creating all valid molecules (with atom type and size limitations) to fill regions of chemical space. These efforts have led to the largest openly available database of 166 billion compounds.[9] Visual maps of chemical space have been created as another way to define, as well as intuitively represent, chemical space. Virshup et al. (2013) used a group of autocorrelation descriptors to define a chemical space, which represents the "small molecular universe", or a space theoretically representative of all possible small molecules < 500 Da. By plotting the contents of PubChem and GDB13 databases on the map created using their descriptors, the authors provided a visualization of the potential sparsity of chemical space.[6] Although they were able to identify gaps and create a library of analogues (via mutations of known molecules) to fill the space around a single compound, the authors did not begin the process of generating molecules to fill in the numerous other gaps.

One way to fill in the gaps of chemical space is through nascent artificial intelligence (AI) approaches. AI models used in cheminformatics can learn the properties, structural information, and even governing laws of chemistry for molecules that they are trained on, to both predict the properties of molecules outside of the training set and to generate potentially novel compounds.[10-12] There have been many recent advancements in the field of AI, for example in artificial neural networks and generative models,[13] which have enabled acceleration of chemical space exploration. An example of one such tool is BioTransformer, a program which uses machine learning (ML), a subfield of AI, to make predictions of small molecule metabolite formation via iterative passage of input structures through metabolic reactions.[14] Another ML model uses a genetic algorithm to generate peptides to fill certain areas of chemical space.[15] Using generative models to produce compounds is not novel in itself and there are many examples in the literature.[12, 16-26] Two deep learning examples of such tools are Mol-CycleGAN,[24] a generative model that creates compounds with a similar structure to the input compound but with optimized property values for increased efficacy against target proteins, and ReLeaSE,[12] a generative model that focuses on building libraries of compounds with narrowly targeted properties.

In this work, our goal was to demonstrate the ability of AI methods to simultaneously fill multiple chemical space gaps without restricting generation to modifying known molecules or for targeting a single (or narrow) chemical property profile. We used DarkChem[17], a variational autoencoder designed to aid in small molecule identification by learning the latent structure of molecules for subsequent property prediction and molecule generation, to target molecular generation for large regions of chemical space, rather than generating compounds that have specific, narrow property values. A number of tools could have been used to generate compounds, but our goal was not to specifically assess the success of the tool itself, but to demonstrate that such a tool could be used to generate compounds with currently undiscovered molecular structures to target large, sparse regions of chemical space. We investigated the unexplored regions of chemical space – a space we defined using ten calculated chemical properties.[17] By generating compounds to fill the gaps in the chemical property space, we gained a better understanding of the architecture of chemical space, as well as demonstrated that AI methods can provide access to previously sparse regions. There is potentially great value in continuing this work specifically to fill chemical space gaps, as in addition to applications in drug discovery and creating representative subset libraries from large compound sets,[6] methods to increase the density and coverage of chemical space could also be used to lower false discovery rates in small molecule identification studies through improving/expanding decoy sets of molecules.

## METHODS

A sample subset of one million compounds was derived from a curated in-house database of 2.5 billion compounds (comprised of metabolites, lipids, drug-like compounds, etc.) which represents the known chemical space of small molecules in publicly available databases such as PubChem, ECMDB, HMDB, DrugBank, ChEMBL, and 35 more. The database was filtered to contain DarkChem-compatible compounds only (i.e. compounds with SMILES of 100 characters or less and containing only carbon, hydrogen, nitrogen, oxygen, phosphorus, and/or sulfur atoms), which left roughly 1.3 billion compounds for sample generation. Compounds were canonicalized, desalted, and neutralized using the open-source cheminformatics software RDKit (v 2019.09.3, rdkit.org),. In order to achieve a representative sample, the subset of one million compounds was chosen in the following way: (1) each of the 40 libraries in the parent database was represented by at least 10 compounds, (2) compounds were sampled according to their mass bin to achieve the same mass distribution as the parent database. This library is available upon request.

Once created, the sample was placed into a previously defined Property Chemical Space (PCS) made up of ten chemical properties using either all compounds (**Figure S1 a**) or only compounds that fell between the $2.5^{th}$ and $97.5^{th}$ percentiles for each axis (**Figure S1 b**).[17] Five properties (ring bond percent, Balaban index, Harary index, LogP, and pKa) were calculated using ChemAxon's *cxcalc* (v 19.13.0), monoisotopic mass was calculated through RDKit, and the remaining four atom ratio properties (N/C, N/H, O/C, O/H) were calculated using a custom Python script. Principal component analysis (PCA) was performed on these 10 properties for the 1 million compounds to create a PCA chemical space; a common method for visualizing chemical space.[6, 7, 15, 17, 27, 28] (**Figure S2**). Properties were then paired together for analysis (ring bond percent and mass, Balaban index and Harary index, LogP and pKa, N/C and N/H, O/C and O/H)

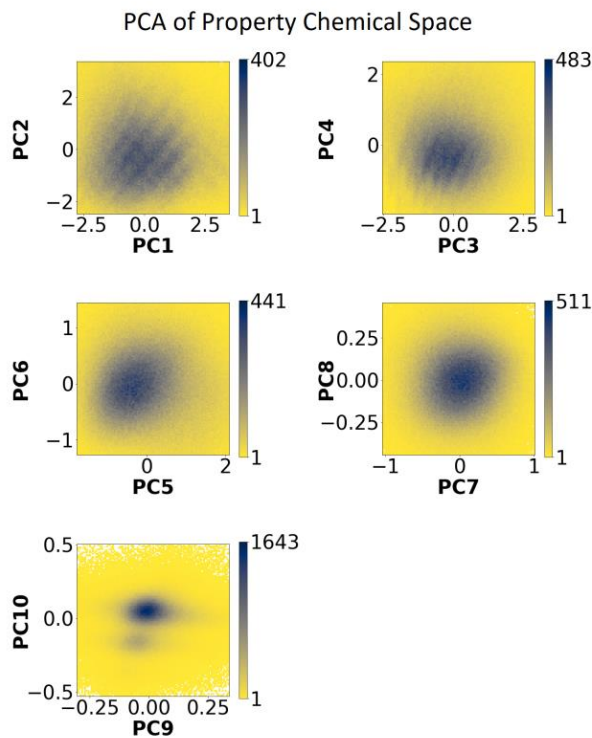

*Figure 1. Chemical space density plots of the data from the compound sample set that fell between the 2.5th and 97.5th percentiles.* *The maximum value of each color bar corresponds to the maximum number of compounds in a single bin for each 2D histogram.*

The sample was also processed by DarkChem to yield 128-dimensional latent vectors.[17] PCA was performed on the latent vectors, and the space was visualized through plots of the first ten principal components. Similar plots to those generated in PCS were also generated in latent space (**Figure S3**).

All plots displaying density of compounds in PCA chemical space (for both PCS and latent space) are 2D histograms with 10,000 bins (100x100). For plotting the sample in PCS, pairwise sets of PCs were used to plot all compounds in PCA space (**Figure S2**) as well as compounds that fell between the 2.5th and 97.5th percentiles for each axis (**Figure 1**). By limiting the axes ranges by percentiles, we hoped to capture roughly 95% of the data and exclude outliers, in order to provide easier visual inspection.

Using the PCA chemical space, we created a hypercube region of interest (ROI). The *numpy* (v 1.18.1) function histogramdd was then used to bin the first three filtered PC dimensions into a 3-dimensional cube comprised of 1 million bins (100x100x100). The ROI was defined from the bins with edges that fell between the range [-1, 1], creating a smaller 30x34x36 cube. This range was chosen based on a desire to find an ROI that would be amenable to testing an AI tool's ability to fill chemical structure gaps. This defined ROI was 68.64% "full" (i.e., 68.64% of bins had at least 1 representative) after mapping the 1 million compounds. The filtering used to create the ROI did not remove compounds from consideration within the ROI area; rather, it simply let us work within a smaller region to find an appropriate ROI.

The compounds in the ROI (88,014 out of the 1 million sample) were used for compound generation. These compounds, represented as SMILES, were encoded into DarkChem latent space vectors and used to define the mean and variance of a 128-dimension normal distribution from which new compounds were sampled. This sampling was repeated using five different variances (1.00, 0.75, 0.50, 0.25, and 0.10) to create the normal distribution to sample molecules from an increasingly ROI-focused region of chemical space.

Because the distribution was created using vectors representing compounds from within the ROI, we hypothesized that the vectors created by randomly sampling this distribution would translate into the majority of novel compounds falling within the desired chemical space ROI, as the output vectors had the same minimum/maximum limits as the input vectors. Generated latent vectors were decoded to SMILES and passed through multiple steps to check for structural and chemical validity. These steps included finding the major tautomer for each molecule using *cxcalc* (cxcalc, 19.23.0, ChemAxon, https://chemaxon.com), neutralizing the major tautomer using ISiCLE,[29] ensuring that the molecules passed through Open Babel,[30] RDKit (v 2019.09.3), and MolVS (v 0.1.1) as valid, and finally removing any molecules with a synthesizability score greater than 6 (on a scale of 1-10 with 10 being hardest to synthesize) as calculated using the SA_Score package from RDKit.

## RESULTS AND DISCUSSION

The 88,014 compounds that make up the ROI in the chemical space defined by applying PCA to all 10 properties at once were used as input for compound generation in DarkChem. Five sampling methods were used to generate compounds: using one standard deviation (calculated from the input compounds) to define the Gaussian distribution, and multiplying the standard deviation by 0.75, 0.5, 0.25, and 0.1. The sampling method that used one standard deviation produced 201,434 valid compounds after filtering. Multiplying the standard deviation by 0.75, 0.5, 0.25, and 0.1 produced 248,488, 214,557, 30,096, and 70 compounds, respectively (Table S1). A list of all of these valid, generated compounds is provided in the Supporting Information. Of the valid compounds generated, 10,035 (4.8%) were found within the ROI for one standard deviation, and 12,663 (5.1%), 8,709 (4.1%), 308 (1%), and 0 (0%) for 0.75, 0.5, 0.25, and 0.1, respectively (**Figure S4, S5**).

We evaluated different methods of sampling to determine if there was a superior way to fill our ROI. We found that using a slightly more narrow Gaussian distribution, created using a standard deviation of 0.75 multiplied by the standard deviation created by the input compounds, was the most successful sampling method for generating novel compounds to fill the gaps in the ROI. As the Gaussian distribution is created from the 88,014 compounds originally in the ROI, it makes sense that a more narrow distribution will produce a larger number of compounds that fall within the ROI because the more narrow distribution means that less sample will be pulled from the tails. Distributions created even more narrow (using 0.5, 0.25, and 0.1) are not as successful as using 0.75 as the distribution is now too narrow and it is possible that it is now less likely to generate a valid molecule (**Table S1**).

When considering all of the compounds generated from the five combined sampling methods, there were 694,645 valid molecules with 31,715 being found in the ROI (**Figure 2**), filling 50.8% of the previously empty voxels in the region (going from 68% full to 85% full). Furthermore, the average number of compounds per voxel in the ROI increased from 2.4 to 3.3, showing that generated compounds not only filled gaps but also increased the chemical space density within the ROI. None of the generated compounds were found in the DarkChem training data or our aggregated 2.5 billion compound database.

Our first attempt at using this AI gap-filling method revealed problems with the validity of the DarkChem generated compounds: radical atoms, pentavalent carbon atoms, and other issues that would prevent the compounds from being synthesized. To minimize these, we instated a more rigorous validity checking pipeline. Through this process, we found that there is a need in the field for a better automated method for checking compound validity, as we had to build our pipeline with a combination of software packages to ensure as many invalid compounds were filtered out as possible. We could not find a singular piece of software which could successfully catch all of the issues of invalidity that we encountered. Another approach for ensuring we are producing valid molecules could be to modify how molecules are represented, for example with SELFIES,[31] where, in theory, every string corresponds to a valid molecule. This could potentially solve the problem of filtering for valid compounds and is something that we will evaluate in the future.

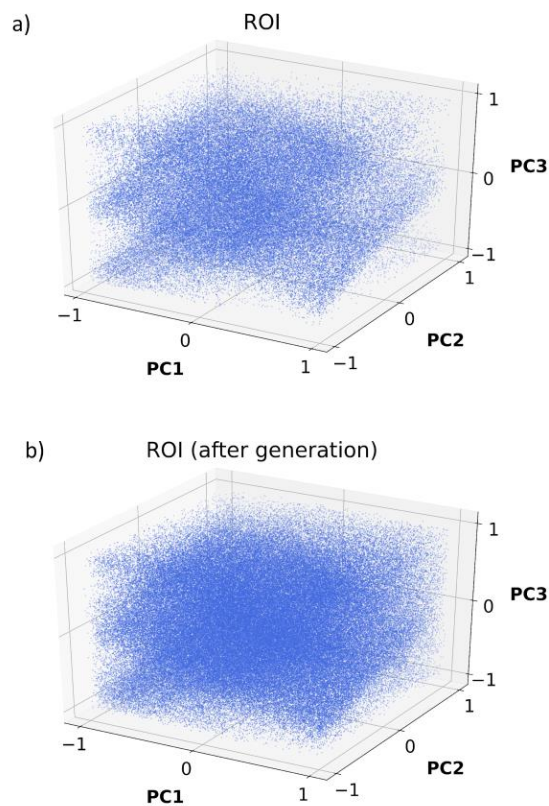

*Figure 2. 3D visualization of ROI before and after being filled with DarkChem generated compounds.* The 88,014 compounds from the base one million sample found in the ROI (a), and the 88,014 compounds from the base one million with the additional 31,715 compounds generated by DarkChem found in the ROI (b).

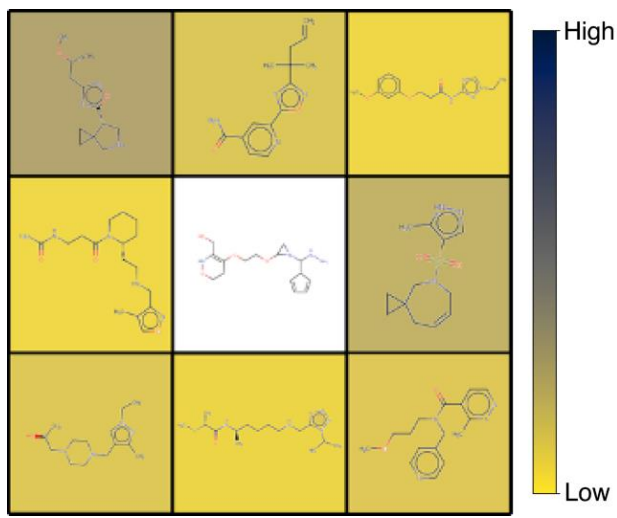

*Figure 3. Example of previously empty voxel filled by DarkChem generation. The middle square of the grid was empty prior to compound generation and holds a novel compound generated by DarkChem. The surrounding squares contained compounds prior to generation, and each square holds a representative compound from the original sample. The background colors represent the number of compounds in each square prior to generation, with white representing the initial void for that region prior to AI compound generation.*

## CONCLUSIONS AND PERSPECTIVES

In this paper, we set out to demonstrate the ability of AI to broadly fill gaps in chemical space using a generative AI model. After selecting an ROI in property chemical space with clear gaps, we used the AI-based tool DarkChem to generate 694,645 novel compounds. These novel compounds not only added to the density of the ROI, they also filled 50.8% of the spaces in the ROI that previously had no compounds.

This initial evaluation of compound generation using the deep-learning tool DarkChem shows promise for exploring and characterizing chemical space. In ensuing work, we plan on exploring optimal approaches for representing chemical space using our full inhouse database of 2.5 billion small molecules. The example here uses a property chemical space as well as a latent chemical space, and one of the questions to be explored is which method(s) are best for chemical space representation: property, AI/learned, or other (e.g. ontological)? The property space explored here is relatively small, with only 10 molecular properties, all calculated from the 2D form of a molecule. Future work will evaluate over one thousand molecular properties, which will include 2D as well as 3D properties. However, we recognize that some of the "gaps", or regions of chemical space where compounds have not yet been identified, may never be filled due to the limits of physics imposed on chemistry. It is our hope that as more compounds are generated through DarkChem and other AI tools, all gaps that can be filled will be occupied, but it is unclear as of now if this is possible with current technology.

Currently, DarkChem does have limitations when it comes to representing chemical space through latent vectors such as only accepting SMILES input, and only SMILES with less than 100 characters made up of carbon, hydrogen, oxygen, sulfur, phosphorus, and nitrogen elements, and the lack of aromatics. These limitations will be addressed with an upcoming version of DarkChem that is currently in development. In addition, we plan to incorporate a different network architecture and larger set of training data.

The complete chemical space of all possible molecules is too large and dense to be explored at once, at least given currently available technology. We believe AI-based tools will continue to provide innovations in the near term that will accelerate the exploration of chemical space beyond what had been possible with traditional cheminformatics/combinatorial-chemistry approaches, with applications for drug discovery, metabolomics, exposomics, and chemical forensics. As technology advances and computer hardware improves to efficiently handle larger amounts of data, brute force methods (e.g. enumeration) could again take a leading role in this work, but for now, AI for compound generation and chemical space exploration appears to be a highly valuable option that will hasten chemical discovery.

## ASSOCIATED CONTENT

## AUTHOR INFORMATION

### Corresponding Authors

Ryan S. Renslow - ryan.renslow@pnnl.gov

Thomas O. Metz - thomas.metz@pnnl.gov

### Author Contributions

The manuscript was written through contributions of all authors. All authors have given approval to the final version of the manuscript.

### Funding Sources

This research was developed with funding from the Defense Advanced Research Projects Agency (DARPA), with additional support provided by the National Institutes of Health, National Institute of Environmental Health Sciences grant U2CES030170. Additional support was provided by the *m/q* Initiative at Pacific Northwest National Laboratory (PNNL). It was conducted under the Laboratory Directed Research and Development Program at PNNL, a multiprogram national laboratory operated by Battelle for the U.S. Department of Energy. PNNL is operated for DOE by Battelle Memorial Institute under contract DE-AC05-76RL01830. The views,